\documentclass[11pt]{article}
\usepackage{hyperref}
\pdfoutput=1
\begin{document}
\title{Landing on the Moon \\ Cratering from a Jet}
\author{Abe Clark, R.P. Behringer \\
\\ Duke University \\ John Brandenburg \\ ORBITEC}
\maketitle
\begin{abstract}
This paper describes work done in investigating cratering processes by impinging jets. It accompanies a video made for the fluid dynamics video presentations at the Division of Fluid Dynamics meeting, November 22-24, 2009.
\end{abstract}

This project characterizes crater formation in a granular
material by a jet of gas impinging on a granular material, such
as a retro-rocket landing on the moon. We
have constructed a 2D model of a planetary surface,
which consists of a thin, clear box partially filled with
granular materials (sand, lunar and Mars simulants...). A metal
pipe connected to a tank of nitrogen gas via a solenoid valve is
inserted into the
top of the box to model the rocket.
The results are recorded using high-speed video. We process
these images and videos in order to test existing models and
develop new ones for describing crater formation. A similar
set-up has been used by Metzger et al (see P. T. Metzger et al. Journal of Aerospace Engineering (2009)). We find that the
long-time shape of the crater is consistent with a predicted
catenary shape (Brandenburg).  The depth and width of the crater
both evolve logarithmically in time, suggesting an analogy to a
description in terms of an activated process: $dD/dt = A
\exp(-aD)$ ($D$ is the crater depth, $a$ and $A$
constants).  This model provides a useful context to understand
the role of the jet speed, as characterized by the pressure used
to drive the flow.  The box width also plays an important role in
setting the width of the crater. Our videos can be found on eCommons here (
\href{http://ecommons.library.cornell.edu/retrieve/59769/Jets-craters-mp1.mpg}{small version}
and
\href{http://ecommons.library.cornell.edu/retrieve/59769/Jets-craters-mp2.mpg}{large version})

\end{document}